\documentclass[a4paper]{article}
\usepackage{theorem}
\theorembodyfont{\upshape}
\newtheorem{theorem}{Theorem}
\newtheorem{definition}{Definition}

\newtheorem{remark}{Remark}

\newtheorem{proposition}{Proposition}

\usepackage{amsmath,amssymb}
\usepackage{mathrsfs}
\usepackage{accents}
\usepackage{graphicx}
\usepackage{cite}
\usepackage[dvipdfmx]{hyperref}
\hypersetup{%
 bookmarksnumbered=true,%
 colorlinks=false,%
 pdftitle={Data Rate Limitations for Stabilization of Uncertain Systems over Lossy Channels},%
 pdfauthor={Kunihisa Okano and Hideaki Ishii},%
 pdfkeywords={Networked control systems; data rate limitations; packet losses; uncertain systems; stabilization}}

\makeatletter
  \def\widebar{\accentset{{\cc@style\underline{\mskip10mu}}}}
  \def\wideubar{\underaccent{{\cc@style\underline{\mskip10mu}}}}
\makeatother

\newcommand{\R}{\mathbb{R}}
\newcommand{\Z}{\mathbb{Z}}
\newcommand{\Zp}{\mathbb{Z}_+}
\newcommand{\N}{\mathbb{N}}
\newcommand{\E}{\textnormal{E}}
\newcommand{\Prob}{\textnormal{Prob}}
\newcommand{\calY}{\mathcal{Y}}
\newcommand{\calA}{\mathcal{A}}
\newcommand{\lambdaP}{\lambda_{A^*}\!}
\newcommand{\lambdaPA}{\lambda_{A}}
\newcommand{\diag}{\textnormal{diag}}
\newcommand{\Rnec}{R_{\textnormal{nec}}}
\newcommand{\bRnec}{\widebar R_{\textnormal{nec}}}

\newcommand{\pnec}{p_{\textnormal{nec}}}

\newcommand{\QED}{\hfill$\blacksquare$\par}
\newcommand{\fref}[1]{Fig.~\ref{#1}}

\newcommand{\TAC}{IEEE Trans.\ Autom.\ Control}
\newcommand{\SCL}{Syst.\ Control Lett.}
\newcommand{\ProcIEEE}{Proc.\ IEEE}
\newcommand{\LNCIS}{Lect.\ Notes Contr.\ Info.\ Sci.}
\newcommand{\SIAM}{SIAM J.\ Control Optim.}
\newcommand{\LA}{Linear Algebra Appl.}

\title{Data Rate Limitations for Stabilization of\\
Uncertain Systems over Lossy Channels}

\author{Kunihisa Okano and Hideaki Ishii
\thanks{The authors are with Department of Computational Intelligence
and Systems Science, Tokyo Institute of Technology, Yokohama, 226-8502, Japan.
Emails: {\tt\small okano@sc.dis.titech.ac.jp}, {\tt\small ishii@dis.titech.ac.jp}.
This work was supported in part by the Ministry of Education, Culture, Sports,
Science and Technology, Japan, under Grant-in-Aid for Scientific Research Grant No.\ 23760385.}%
}

\date{January 5th, 2012}

\begin{document}

\maketitle

\begin{abstract}
This paper considers data rate limitations for mean square stabilization of
uncertain discrete-time linear systems via finite data rate and lossy channels.
For a plant having parametric uncertainties,
a necessary condition and a sufficient condition are derived,
represented by the data rate, the packet loss probability,
uncertainty bounds on plant parameters,
and the unstable eigenvalues of the plant.
The results extend those existing in the area of
networked control, and in particular, the condition is exact
for the scalar plant case.
\end{abstract}

\section{Introduction}
Networked control systems have attracted 
much research interest in recent years \cite{Antsaklis2007, Bemporad2010}.
In such systems,
the communication between the plant and the controller is constrained
due to the use of shared channels.
Though modern communication channels
may have large bandwidth,
in the systems consisting of sensors, actuators, and other devices,
each component can be allocated only a portion of it
for the required real-time transmissions.
Thus, it is important to study fundamental
relations concerning the control performance and the communication constraints
in such systems.

One such relation is the limitation on the {\it data rate}
for the stabilization of unstable linear systems.
The seminal works of \cite{Wong1999, Nair2004, Tatikonda2004}
have presented the minimum data rate for stabilization,
and have shown that it only depends on the product of the unstable eigenvalues of the plant.
Another fundamental relation can be found under the presence of {\it packet losses}.
In practical channels,
transmitted packets may be lost
due to congestion or delay.
In \cite{Elia2005},
it is shown that
the maximum loss probability tolerable for mean square stabilization
is described also by the product of the unstable eigenvalues of the plant.
It is worth noticing that both limitations require more communication for plants
with more unstable dynamics.
On the other hand, the works \cite{You2010, Tatikonda2004a} focus on
the two constraints of data rates and packet losses simultaneously
and extend the results on limitations.
It is interesting that while the work \cite{Tatikonda2004a} studies
the notion of stability with probability 1,
the paper \cite{You2010} is concerned with mean square stability.
As a consequence, different bounds are obtained.
Furthermore, various communication constraints are considered
(e.g., delays, variable sampling periods, sharing the channel among multiple nodes),
not only for linear systems but also for nonlinear systems
(see, e.g., \cite{Liberzon2005, Nesic2009, Heemels2010}).

While in the existing works,
it is commonly assumed that the exact plant model is known,
when the plant is uncertain,
fewer results are available.
The work of \cite{Phat2004} deals with
a linear time-invariant plant with norm-bounded uncertainty;
and in \cite{Martins2006}, a nonlinear scalar plant
with stochastic uncertainty is considered.
Both papers give sufficient conditions on the data rate
for stabilization using a lossless channel.
However,
these limitations may contain some conservativeness,
and it is still unclear at least
how large the data rate should be when the plant is uncertain.

The objective of this paper is to provide an answer to this question.
We consider plants with parametric uncertainty.
Though this class of uncertainty is different from those in the works mentioned above,
it is widely adopted
and various results are established (see \cite{Barmish1994} and references therein).
To deal with the uncertainty,
we follow the approach of \cite{You2010},
where
the plants are given in the controllable canonical forms.
Hence, the uncertainty can be considered as a natural extension of this approach.
As to the communication constraints,
following the setup in \cite{You2010},
we consider the channel is with data rate constraints and packet losses.

We derive a necessary condition and a sufficient condition,
which provide
bounds on the required data rate and loss probability
to stabilize the closed-loop system.
The results have several features:
First, the limitation given by the conditions is tight
for the scalar plants case, which is not attained in \cite{Phat2004, Martins2006}.
Second, it generalizes the results in \cite{You2010} in the sense
that without uncertainty in the plants,
the bounds coincide with that in \cite{You2010}.
Moreover, it is known that for plants without uncertainty,
stabilization is possible with any data rate greater than the necessary bound
by introducing a time-sharing protocol \cite{Nair2004, Tatikonda2004, You2010}.
However, we show that this may be difficult in the uncertain case.

The paper is organized as follows.
In Section \ref{formulation},
we describe the setup of the networked control systems considered.
Then, we present a necessary condition and a sufficient condition
in Sections \ref{necessary} and \ref{sufficient}, respectively.
The limitation for the system employing a time-sharing protocol is
discussed in Section \ref{average},
and finally we provide concluding remarks in Section \ref{conclusion}.

\section{Problem Formulation}\label{formulation}
We consider the stabilization of a networked control system
which includes a digital channel
between the plant and the controller.
At time $k\in\Zp$, the encoder observes the plant output $y_k\in\R$,
and quantizes it as an $R$-bit signal $s_k\in\Sigma^R$, where $R\geq0$.
Here, the set $\Sigma^R$ represents all possible outputs of the encoder,
and contains $2^R$ symbols.
The quantized signal $s_k$ is transmitted to the decoder via a lossy channel.
Based on the received signal,
the decoder generates the interval $\calY_k\subset\R$,
which is an estimate of the plant output $y_k$.
Finally, the controller receives this set $\calY_k$
and provides the control input $u_k\in\R$.
\fref{fig,system} shows the connections among components of the system.
\begin{figure}[t]
\begin{center}
\unitlength 0.1in
\begin{picture}( 29.0000, 14.1000)(  9.8000,-18.0000)
%
\special{pn 8}%
\special{pa 1000 400}%
\special{pa 1700 400}%
\special{pa 1700 800}%
\special{pa 1000 800}%
\special{pa 1000 400}%
\special{pa 1700 400}%
\special{fp}%
\put(12.0000,-6.6000){\makebox(0,0)[lb]{Plant}}%
%
\special{pn 8}%
\special{pa 2200 400}%
\special{pa 2900 400}%
\special{pa 2900 800}%
\special{pa 2200 800}%
\special{pa 2200 400}%
\special{pa 2900 400}%
\special{fp}%
\put(23.1000,-6.6000){\makebox(0,0)[lb]{Encoder}}%
%
\special{pn 8}%
\special{pa 2200 1400}%
\special{pa 2900 1400}%
\special{pa 2900 1800}%
\special{pa 2200 1800}%
\special{pa 2200 1400}%
\special{pa 2900 1400}%
\special{fp}%
\put(23.2000,-16.5500){\makebox(0,0)[lb]{Decoder}}%
%
\special{pn 8}%
\special{pa 1000 1400}%
\special{pa 1700 1400}%
\special{pa 1700 1800}%
\special{pa 1000 1800}%
\special{pa 1000 1400}%
\special{pa 1700 1400}%
\special{fp}%
\put(10.6000,-16.6000){\makebox(0,0)[lb]{Controller}}%
%
\special{pn 8}%
\special{pa 1700 600}%
\special{pa 2190 600}%
\special{fp}%
\special{sh 1}%
\special{pa 2190 600}%
\special{pa 2124 580}%
\special{pa 2138 600}%
\special{pa 2124 620}%
\special{pa 2190 600}%
\special{fp}%
%
\special{pn 8}%
\special{pa 2200 1600}%
\special{pa 1700 1600}%
\special{fp}%
\special{sh 1}%
\special{pa 1700 1600}%
\special{pa 1768 1620}%
\special{pa 1754 1600}%
\special{pa 1768 1580}%
\special{pa 1700 1600}%
\special{fp}%
%
\special{pn 8}%
\special{pa 1350 1400}%
\special{pa 1350 800}%
\special{fp}%
\special{sh 1}%
\special{pa 1350 800}%
\special{pa 1330 868}%
\special{pa 1350 854}%
\special{pa 1370 868}%
\special{pa 1350 800}%
\special{fp}%
\put(18.7000,-5.5000){\makebox(0,0)[lb]{$y_k$}}%
\put(31.5500,-5.5000){\makebox(0,0)[lb]{$s_k$}}%
\put(31.5000,-15.5000){\makebox(0,0)[lb]{$\gamma_{k}s_k$}}%
\put(25.1000,-11.7000){\makebox(0,0)[rb]{$\gamma_{k-1}$}}%
\put(13.2000,-11.7000){\makebox(0,0)[rb]{$u_k$}}%
%
\special{pn 8}%
\special{pa 2550 1400}%
\special{pa 2550 800}%
\special{fp}%
\special{sh 1}%
\special{pa 2550 800}%
\special{pa 2530 868}%
\special{pa 2550 854}%
\special{pa 2570 868}%
\special{pa 2550 800}%
\special{fp}%
\put(18.7000,-15.5000){\makebox(0,0)[lb]{$\mathcal{Y}_k$}}%
\put(34.4500,-10.9500){\makebox(0,0)[lb]{Lossy}}%
\put(33.7000,-11.7500){\makebox(0,0)[lb]{channel}}%
%
\special{pn 8}%
\special{pa 3600 606}%
\special{pa 3600 920}%
\special{fp}%
\special{sh 1}%
\special{pa 3600 920}%
\special{pa 3620 854}%
\special{pa 3600 868}%
\special{pa 3580 854}%
\special{pa 3600 920}%
\special{fp}%
%
\special{pn 8}%
\special{pa 3600 1290}%
\special{pa 3600 1600}%
\special{fp}%
%
\special{pn 8}%
\special{ar 3600 1100 280 186  0.0000000 6.2831853}%
%
\special{pn 8}%
\special{pa 2900 600}%
\special{pa 3600 600}%
\special{fp}%
%
\special{pn 8}%
\special{pa 3600 1600}%
\special{pa 2900 1600}%
\special{fp}%
\special{sh 1}%
\special{pa 2900 1600}%
\special{pa 2968 1620}%
\special{pa 2954 1600}%
\special{pa 2968 1580}%
\special{pa 2900 1600}%
\special{fp}%
\end{picture}%
\caption{Networked control system}
\label{fig,system}
\end{center}
\end{figure}
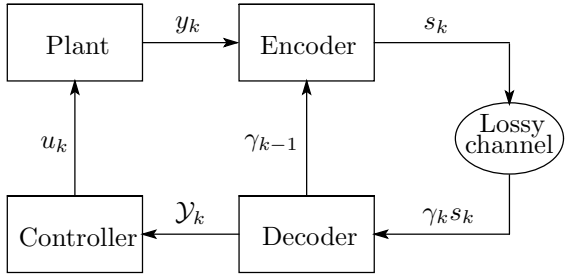

Now we describe the details of each element in the system.
The plant is an $n$-dimensional single-input single-output
autoregressive system whose parameters have uncertainties
and may be time varying:
\begin{align}
  y_{k+1}-a_{1,k}y_k-a_{2,k}y_{k-1}-\cdots-a_{n,k}y_{k-n+1}=u_k.\label{AR}
\end{align}
Here, the initial value $y_0$ of the output is bounded with a known interval $|y_0|\leq Y_0$,
and $y_k=0$ for $k<0$.
Each uncertain parameter $a_{i,k} $ is represented by
the nominal value $a_i^*$
and the width $\epsilon_i\!\geq\!0$ of the perturbation as
\begin{gather}
 a_{i,k}\!\in\!\calA_i\!:=\!\left[a_i^*-\epsilon_i, a_i^*+\epsilon_i\right]\;
 \text{for }i=1,2,\dots,n,\label{uncertainty}
\end{gather}
where $\calA_i$ represents the uncertainty of the $i$th parameter $a_{i,k}$.

The plant (\ref{AR}) can be rewritten in the controllable canonical form as
\begin{align}
 x_{k+1}=A_kx_k+Bu_k,\;
 y_{k}=Cx_k,\label{LTI}
\end{align}
where $x_k:=[y_{k-n+1}\;y_{k-n+2}\;\cdots\;y_k]^T$ and
\begin{align}
 A_k=&
 \left[\begin{array}{cccc}
  0& 1&  \cdots& 0\\
  \vdots & \ddots& \ddots& \vdots\\
  0& 0& \cdots&1\\
  a_{n,k}& a_{n-1,k}& \cdots&a_{1,k}
 \end{array}\right]\in\R^{n\times n},\
 B=
 \left[
  \begin{array}{c}
   0\\
   \vdots\\
   0\\
   1
 \end{array}\right]\in\R^n,\notag\\
 C=&
 \left[
 \begin{array}{cccc}
  0& \cdots& 0& 1\\
 \end{array}\right]\in\R^{1\times n}.\notag
\end{align}
Let $A^*$ denote the nominal $A$-matrix,
and let $\lambdaP$ be the product of the eigenvalues of $A^*$:
\begin{align}
 A^*:=\left[\begin{array}{cccc}
 0& 1& \cdots & 0\\
 \vdots & \ddots& \ddots& \vdots\\
 0& 0& \cdots&1\\
 a_n^*& a_{n-1}^*& \cdots&a_1^*
\end{array}\right],\
\lambdaP:=\prod_{i=1}^n\lambda_i(A^*)=a_n^*,\notag
\end{align}
where $\lambda_i(\cdot)$ represents an eigenvalue of a matrix.
Assume that all eigenvalues of the matrix $A_k$ are unstable
for the uncertain parameters in (\ref{uncertainty}).
In particular, this implies
\begin{align}
  |a_n^*|-\epsilon_n>1.\label{a_n^*ineq}
\end{align}

The encoder $E_k$ at time $k$
quantizes the output $y_k\in\R$ to the $R$-bit signal $s_k\in\Sigma^R$,
and then transmits it to the decoder via a lossy channel.
Let the random variable $\gamma_k$ represent the state of packet reception/loss
at time $k$.
The encoder knows the state of the previous step, $\gamma_{k-1}$,
through the acknowledgement signal from the decoder.

In order to deal with the uncertainty in the plant,
we introduce some structure in the encoder.
Specifically, it is realized by a finite level uniform quantizer.
Such a class of encoders has been employed in,
e.g., \cite{Tatikonda2004, Phat2004, Martins2006, You2010}.
We will have more discussion later.
Let $q_N(\cdot)$ denote the $N$-level uniform quantizer with $N=\log_2 R$
and the input range $\left[-1/2,1/2\right]$.
It partitions the range $\left[-1/2,1/2\right]$
into $N$ intervals of widths $1/N$ as
\begin{align}
 q_N(y):=\begin{cases}
	 i & \text{if } -\frac{1}{2}+\frac{i}{N}\leq y < -\frac{1}{2}+\frac{i+1}{N},\
	  i\in\{0,\,1,\dots,N-2\},\\
	 N-1 & \text{if } \frac{1}{2}-\frac{1}{N}\leq y \leq \frac{1}{2}.
	\end{cases}\notag
\end{align}
Let the set of its outputs be $\Sigma^R:=\{0,1,\dots,N-1\}$.
Then the encoder $E_k$ is given by
\begin{align}
 s_k
 =E_k(y_0^{k},\gamma_{0}^{k-1})
 =q_{N}\left(\frac{y_k}{\sigma_k}\right),\notag
\end{align}
where $y_0^k$ is the sequence $y_0^{k}:=\left\{y_0,\dots,y_k\right\}$
and similarly $\gamma_{0}^{k-1}:=\left\{\gamma_0,\dots,\gamma_{k-1}\right\}$.
We use similar notations in the sequel.
In addition,
$\sigma_k>0$ is the scaling parameter of the input range.
The above equation implies that the input range of the encoder
is $\left[-\sigma_k/2,\sigma_k/2\right]$.
The scaling parameter $\sigma_k$ must be computable on both sides of the channel.
Since the receiver side knows only $\gamma_k s_k$,
this parameter must be updated as
\begin{align}
  \sigma_k=S_k((\gamma s)_0^{k-1}),\;
 \sigma_0\geq Y_0,\notag
\end{align}
where $(\gamma s)_k:=\gamma_k s_k$.

The transmitted signal $s_k$ is randomly lost
due to unreliability in the channel caused by congestion or delay.
The packet reception/loss state at time $k$ is represented by
the random variable $\gamma_k\in\{0,1\}$.
If $\gamma_k=0$ then the packet is lost at time $k$;
otherwise, it arrives successfully.
The process $\{\gamma_k\}_{k=0}^\infty$ is independent and
identically distributed (i.i.d.) with the loss probability
specified by $p\in[0,1)$, i.e., $\Prob\left(\{\gamma_k=0\}\right)=p$.

The decoder $D_k$ converts the received signal $(\gamma s)_k$ to the interval
$\calY_k\subset[-\sigma_k/2,\sigma_k/2]$ by
$ \calY_k=D_k\left((\gamma s)_0^{k},\gamma_0^k\right)$.
The interval $\calY_k$ is an estimation set that the plant output
$y_k$ should be included in.
If the packet $s_k$ arrives, i.e., if $\gamma_k=1$,
then the output interval $\calY_k$ corresponds to one of the partitions of the quantizer;
otherwise, $\calY_k$ is equal to the entire input range of the encoder.
Hence, we have
\begin{gather}
 \calY_k=
 \begin{cases}
  \left[\left.-\frac{\sigma_k}{2}+\frac{\sigma_k}{N}i,-\frac{\sigma_k}{2}+\frac{\sigma_k}{N}(i+1)\right)\right.
   & \text{if }\gamma_k=1 \text{ and } s_k=i,\ i\in\left\{0,1,\dots,N-2\right\},\\
  \left[\frac{\sigma_k}{2}-\frac{\sigma_k}{N},\frac{\sigma_k}{2}\right]
   & \text{if }\gamma_k=1 \text{ and } s_k=N-1,\\
  \left[-\frac{\sigma_k}{2},\frac{\sigma_k}{2}\right]
   & \text{if } \gamma_k=0.
 \end{cases}\label{def,calY}
\end{gather}

The controller $C_k$ generates the control input $u_k$
based on the estimation set $\calY_k$ as $u_k=C_k(\calY_{0}^k)$.

It is remarked that
the scaling parameter $\sigma_k$ should be large enough
to cover all possible inputs to the encoder.
Otherwise,
the quantizer may be saturated, in which case we lose track of the plant output $y_k$.
On the other hand,
if we take $\sigma_k$ large, the quantization error also  becomes large.
Moreover, to achieve stabilization of the system $\sigma_k$ should decay gradually.

We determine the scaling parameter as follows.
At time $k$,
the encoder and the decoder predict the next plant output $y_{k+1}$
based on the observed $\calY_0^k$.
Let $\calY_{k+1}^-\subset\R$ be
the set of all possible outputs $y_{k+1}$ of the uncertain system (\ref{AR}).
Then the scaling parameter $\sigma_{k+1}>0$ is chosen such that
\begin{align}
 \sigma_{k+1}\geq\mu(\calY_{k+1}^-),\label{sigma_ineq}
\end{align}
where $\mu(\cdot)$ denotes the Lebesgue measure on $\R$.

As the prediction set $\calY_{k+1}^-$,
we employ the following one.
\begin{definition}\label{def,estimated_state_set}
 The prediction set of the plant output $y_{k+1}$ constructed at time $k$
is defined as follows:
 \begin{align}
  \calY^-_{k+1}:=\{&a_1'y_k'+\cdots+a_n'y_{k-n+1}' :
  a_1'\in\calA_1,\dots, a_n'\in\calA_n,\
  y_k'\in\calY_{k}, \dots, y_{k-n+1}'\in\calY_{k-n+1}
  \}.\label{def,calY^-}
 \end{align}
\end{definition}

Under this definition,
our prediction strategy is to use the information
regarding $y_k,\dots,y_{k-n+1}$ independently such that
$y_{k-i+1}\in\calY_{k-i+1}$ for each $i=1,2,\dots,n$,
where $\calY_{k-i+1}$ is the interval received on the decoder side at time $k-i+1$.
Then, clearly, $\mu(\calY_{k+1}^-)$ is large enough to include $y_{k+1}$,
and it is computable on both sides of the channel.

\begin{remark}
 The class of prediction sets in Definition \ref{def,estimated_state_set} is
employed to pursue an analytical approach.
There may be other prediction methods that generate
a less conservative prediction set $\widetilde\calY_{k+1}^{-}$
satisfying $\mu(\widetilde\calY_{k+1}^{-})<\mu(\calY_{k+1}^-)$.
If, for example, we take $y_k,\dots,y_{k-n+1}$ not independently
but by looking at the correlations among them,
then the estimation sets $\calY_{k-1},\dots,\calY_{k-n+1}$
from times before $k$ and $\calY_{k+1}^{-}$ may shrink.
In the case of uncertain plants, however,
it is difficult to analytically minimize the Lebesgue measures
of the prediction sets containing $y_{k+1}$,
though it may be possible numerically \cite{Rohn1989}.
\end{remark}

The control objective is stabilizing the system
depicted in \fref{fig,system}
in a stochastic sense as described below.
\begin{definition}
 The system depicted in \fref{fig,system} is mean square stable (MSS)
if the plant output $y_k$ asymptotically goes to zero in the mean square sense
for all possible uncertainties within the bounds in (\ref{uncertainty}),
i.e.,
\begin{align}
 \lim_{k\to\infty}\E[|y_{k}|^2]=0,\
 \forall a_{1,k}\in\calA_1,\forall a_{2,k}\in\calA_2, \dots,\forall a_{n,k}\in\calA_n,\;
 \forall k\in\Zp.\notag
\end{align}
\end{definition}

The problem of the paper is to
find limitations on the data rate and the loss probability
for the system to be MSS.

\section{Necessary Condition}\label{necessary}
In this section, we present the first main result of the paper:
A necessary condition for the system to be MSS.
We introduce two notations representing certain lower bounds on the data rate:
\begin{align}
 \Rnec^{(0)}&:=
 \log_2\frac{\left(|\lambdaP|+\epsilon_n\right)\sqrt{1-p}}{\sqrt{1-p(|\lambdaP|+\epsilon_n)^2}},\notag\\
 \Rnec^{(1)}&:=
 \log_2\frac{\left(|\lambdaP|-\epsilon_n\right)\sqrt{1-p}}{\sqrt{1-p(|\lambdaP|+\epsilon_n)^2}-\epsilon_n\sqrt{1-p}}.\notag
\end{align}

The following theorem is the necessity result.
\begin{theorem}\label{th,nec}
 If the system depicted in \fref{fig,system} is MSS,
then the following inequalities hold:
\begin{align}
 R&>\Rnec:=\max\{\Rnec^{(0)},\Rnec^{(1)}\},\label{nec,R}\\
 p&<\pnec:=\frac{1-\epsilon_n^2}{(|\lambdaP|+\epsilon_n)^2-\epsilon_n^2},\
 0\leq \epsilon_n<1.\label{nec,p}
\end{align}
\end{theorem}

One can verify that $\Rnec$
is monotonically increasing with respect to
$|\lambdaP|$, $p$, and $\epsilon_n$ (under the constraints of (\ref{nec,p})),
and similarly $\pnec$ is monotonically decreasing.
In particular, as expected, more uncertainty in the plant will result in
higher requirement in communication with a larger data rate
and a smaller loss probability.

A special case of this result is when there is no uncertainty in the plant,
in which case we have $\epsilon_n=0$.
Then, the bounds in the theorem coincide with
the ones given in \cite{You2010}.
\begin{proposition}[\!\!\cite{You2010}]\label{prop,you}
 Suppose that in (\ref{LTI}) the exact plant model is known
and is time-invariant,
represented by the triple $(A,B,C)$.
Then, the system in \fref{fig,system} is MSS if and only if the data rate
and the loss probability satisfy the following inequalities:
\begin{align}
 R&>R_{\text{Y}}(\lambdaPA):=\log_2
 \frac{|\lambdaPA|\sqrt{1-p}}{\sqrt{1-p|\lambdaPA|^2}},\label{you_R}\\
 p&<p_{\text{Y}}(\lambdaPA):=\frac{1}{|\lambdaPA|^2}.\label{you_p}
\end{align}
Here, $\lambdaPA$ denotes the product of the eigenvalues of $A$,
that is,
$\lambdaPA:=\prod_{i=1}^n\lambda_i(A)$.
\end{proposition}

It is clear that when $\epsilon_n=0$, then 
we have that $\Rnec=R_{\text{Y}}(\lambdaP)$ and $\pnec=p_{\text{Y}}(\lambdaP)$.
Thus, Theorem \ref{th,nec} includes Proposition \ref{prop,you} as a special case.
For the case $\epsilon_n>0$,
we have the following inequalities
\begin{align}
 \Rnec&\geq\max_{\lambdaPA\in\calA_n}R_{\text{Y}}(\lambdaPA)
 =\Rnec^{(0)},\notag\\
 \pnec&<\min_{\lambdaPA\in\calA_n}p_{\text{Y}}(\lambdaPA).\notag
\end{align}
By these relations, 
we have that when the plant is uncertain with $\epsilon_n>0$,
then even if we assume the most conservative plant dynamics,
the limitations $R_{\text{Y}}$ and $p_{\text{Y}}$
given by Proposition \ref{prop,you}
may not satisfy the necessary conditions (\ref{nec,R}) and (\ref{nec,p}).

In Theorem \ref{th,nec},
the condition $\epsilon_n<1$ indicates the existence of a threshold
on the uncertainty bound, over which the system cannot be stabilized
with any controller regardless of the size of the data rate in the channel.

Before starting the proof of Theorem \ref{th,nec},
we introduce an inequality regarding the set $\calY_{k+1}^-$.
By the definition in (\ref{def,calY^-}),
the prediction set $\calY_{k+1}^-$ satisfies
\begin{align}
 \mu(\calY^-_{k+1})
 =\sum_{i=1}^n\mu\left(\calA_i\calY_{k-i+1}\right)
 \geq\mu\left(\calA_n\calY_{k-n+1}\right)\label{BM}
\end{align}
where
\begin{align}
 \calA_i\calY_{k-i+1}
 :=\left\{a_i'y'_{k-i+1} : a_i'\in\calA_i,\,y_{k-i+1}'\in\calY_{k-i+1}\right\}\notag
\end{align}
for $i=1,2,\dots,n$.
Here, the equality holds by the Brunn-Minkowski inequality \cite{Cover2006}.

{\it Proof of Theorem \ref{th,nec}:~}
First, we show that the mean square stability of the plant output $y_k$
implies that the scaling parameter $\sigma_k$ is also MSS.
For any measurable set $\calY_k\subset\R$,
it is obvious that
\begin{align}
 2\max_{y_k'\in\calY_k}|y_k'|\geq\mu(\calY_k).\notag
\end{align}
By the definition (\ref{def,calY}) of the decoder,
the right-hand side of the inequality satisfies
$\mu(\calY_k)=\sigma_k/N^{\gamma_k}$.
Hence, if $\lim_{k\to\infty}\E[|y_k|^2]=0$,
then $\lim_{k\to\infty}\E[\sigma_k^2]=0$ holds.

In the rest of the proof,
we study a necessary condition for $\sigma_k$ to be MSS.
Notice from (\ref{sigma_ineq}) that $\sigma_{k+1}$ is bounded
from below by $\mu(\calY_{k+1}^-)$.
Substitution of (\ref{BM}) into (\ref{sigma_ineq}) yields
\begin{align}
 \sigma_{k+1}\geq\mu\left(\calA_n\calY_{k-n+1}\right).\label{nec,0}
\end{align}

We next evaluate the right-hand side of the above inequality.
This term depends on where the intervals  $\calA_n$ and $\calY_{k-n+1}$
lie with respect to the origin of the real axis.
By (\ref{uncertainty}), we have $\calA_n=[a_n^*-\epsilon_n,a_n^*+\epsilon_n]$,
but (\ref{a_n^*ineq}) implies that it does not contain the origin.
We first consider the case $a_n^*>0$.
Using basic notions from interval arithmetics \cite{Moore1966},
we obtain
\begin{align}
 \mu\left(\calA_n\calY_{k-n+1}\right)
 =a_n^*\mu(\calY_{k-n+1})+\epsilon_n\beta(\calY_{k-n+1}),\label{nec,1a}
\end{align}
where $\beta(\cdot)$ is given by
\begin{gather}
 \beta(\calY_k):=
 \begin{cases}
  \widebar\calY_k+\wideubar\calY_k  & \text{if } 0\leq\wideubar\calY_k,\\
  \widebar\calY_k-\wideubar\calY_k  & \text{if } \wideubar\calY_k<0<\widebar\calY_k,\\
  -\widebar\calY_k-\wideubar\calY_k  & \text{if } \widebar\calY_k\leq0,
 \end{cases}\label{def,beta}\\
 \widebar\calY_k:=\sup_{y_k'\in\calY_k}y_k',\;
 \wideubar\calY_k:=\inf_{y_k'\in\calY_k}y_k'.\notag
\end{gather}
Similarly, for the case $a_n^*<0$, we have
\begin{align}
 \mu\left(\calA_n\calY_{k-n+1}\right)
 =-a_n^*\mu(\calY_{k-n+1})+\epsilon_n\beta(\calY_{k-n+1}).\label{nec,1b}
\end{align}
Clearly, we can write (\ref{nec,1a}) and (\ref{nec,1b}) in one form:
\begin{align}
 \mu\left(\calA_n\calY_{k-n+1}\right)
 =|a_n^*|\mu(\calY_{k-n+1})+\epsilon_n\beta(\calY_{k-n+1}).\label{nec,1}
\end{align}
Though the value on the right-hand side of (\ref{nec,1}) may vary with $\calY_{k-n+1}$,
the inequality (\ref{nec,0}) holds for any $\calY_{k-n+1}$.

We claim that the maximum of $\mu\left(\calA_n\calY_{k-n+1}\right)$ in (\ref{nec,1})
over all possible intervals $\calY_{k-n+1}$ at time $k-n+1$ can be written as
\begin{gather}
\max_{\calY_{k-n+1}}\mu\left(\calA_n\calY_{k-n+1}\right)
 =\eta_k\sigma_{k-n+1},\label{nec,betamax}
\end{gather}
where $\eta_k$ is the random variable given by
\begin{gather}
 \eta_k:=\frac{|a_n^*|+\max\left\{N^{\gamma_{k-n+1}}-1,1\right\}\epsilon_n}
 {N^{\gamma_{k-n+1}}}.\label{def,eta}
\end{gather}
In (\ref{nec,1}),
the first term is $|a_n^*|\mu(\calY_{k-n+1})
=|a_n^*|\sigma_{k-n+1}/N^{\gamma_{k-n+1}}$.
For the second term, we must consider the following three cases.

{\bf (i)} $0\leq\wideubar\calY_{k-n+1}$:
In this case, by (\ref{def,calY}),
it is necessary that $\gamma_{k-n+1}=1$ and $N\geq 2$.
From (\ref{def,calY}) and (\ref{def,beta}), we have
\begin{align}
 \max_{\calY_{k-n+1}}\beta(\calY_{k-n+1})
 =\frac{N-1}{N}\sigma_{k-n+1}.\label{beta,case1}
\end{align}

{\bf (ii)} $\wideubar\calY_{k-n+1}<0<\widebar\calY_{k-n+1}$:
If $\gamma_{k-n+1}=0$ or $N<2$, then this condition is satisfied for any $\calY_{k-n+1}$.
By (\ref{def,beta}), we obtain $\beta(\calY_{k-n+1})= \widebar\calY_{k-n+1}-\wideubar\calY_{k-n+1}
=\mu(\calY_{k-n+1})$.
Hence, we have
\begin{align}
 \max_{\calY_{k-n+1}}\beta(\calY_{k-n+1})
=\frac{\sigma_{k-n+1}}{N^{\gamma_{k-n+1}}}.\notag
\end{align}

{\bf (iii)} $\widebar\calY_{k-n+1}\leq 0$:
This case can be reduced to {\bf (i)}.
We hence obtain (\ref{beta,case1}).

From the above, the relation in (\ref{nec,betamax}) is derived.

By (\ref{nec,0}) and (\ref{nec,betamax}), it holds that
\begin{align}
 \sigma_{k+1}\geq\eta_k\sigma_{k-n+1}.\label{sigma_recurrence}
\end{align}
This is a key inequality, which is a consequence of $\sigma_k$ being MSS.
Noticing that $\sigma_k>0$ for each time $k$,
take the square of both sides of (\ref{sigma_recurrence}) as
\begin{align}
 \E[\sigma_k^2]
 \geq\E[\eta_{k}^2\sigma_{k-n+1}^2]=\E[\eta_{k}^2]\E[\sigma_{k-n+1}^2].\notag
\end{align}
Here, the equality holds due to the independence of
$\sigma_{k-n+1}$ and $\gamma_{k-n+1}$;
note that $\sigma_{k-n+1}$ is an element of the $\sigma$-field generated by
the sequence $\gamma_0^{k-n}$.
The above inequality shows that if $\sigma_k$ is MSS
then $\eta_k$ must satisfy $\E[\eta_{k}^2]<1$.

The next step is to obtain a lower bound on $N$
from the inequality $\E[\eta_k^2]<1$.
For this, we must consider two cases.

{\bf (i)} $1\leq N<2$:
By the definition of $\eta_k$ in (\ref{def,eta}), in this case
we have $\eta_k=(|a_n^*|+\epsilon_n)/N^{\gamma_{k-n+1}}$.
Hence, it holds that
\begin{align}
 &\E[\eta_{k}^2]
 =p(|a_n^*|+\epsilon_n)^2+(1-p)\left(\frac{|a_n^*|+\epsilon_n}{N}\right)^2<1\notag\\
 &\Leftrightarrow
 N>\frac{(|a^*_n|+\epsilon_n)\sqrt{1-p}}{\sqrt{1-p(|a^*_n|+\epsilon_n)^2}},\
 p<\pnec^{(0)}:=\frac{1}{(|a_n^*|+\epsilon_n)^2}.\notag
\end{align}
Using $R=\log_2 N$, we can simplify this as
\begin{align}
 R>\Rnec^{(0)},\ p<\pnec^{(0)}.\label{nec,3,R0}
\end{align}

{\bf (ii)} $N\geq 2$:
As in {\bf (i)}, $\E[\eta_{k}^2]<1$ is equivalent to
\begin{align}
 R>\Rnec^{(1)},\ p<\pnec^{(1)}:=\frac{1-\epsilon_n^2}{|a_n^*|^2+2|a_n^*|\epsilon_n}.\label{nec,3,R1}
\end{align}

We next simplify these necessary conditions (\ref{nec,3,R0}) and (\ref{nec,3,R1})
into one form.
First, we consider the constraints on the data rate $R$.
According to (\ref{nec,3,R0}) and (\ref{nec,3,R1}),
a lower bound on $R$ is $\Rnec^{(0)}$ if $0\leq R<1$, and $\Rnec^{(1)}$ otherwise.
However, by the definitions of $\Rnec^{(0)}$ and $\Rnec^{(1)}$,
we can show the three relations
\begin{align}
  \Rnec^{(0)}>1 &\Rightarrow \Rnec^{(1)}> \Rnec^{(0)},\notag\\
  \Rnec^{(0)}<1 &\Rightarrow \Rnec^{(1)}< \Rnec^{(0)},\notag\\
  \Rnec^{(0)}=\Rnec^{(1)} &\Rightarrow \Rnec^{(0)}=\Rnec^{(1)}=1.\notag
\end{align}
That is, the bound $\Rnec^{(1)}$ is larger than $\Rnec^{(0)}$ when $\Rnec^{(0)}>1$,
is smaller when $\Rnec^{(0)}<1$,
and crosses $\Rnec^{(0)}$ when $\Rnec^{(1)}=1$.
Thus, the constraints on $R$ can be reduced to the desired bound
$ R>\max\{\Rnec^{(0)},\Rnec^{(1)}\}$ in (\ref{nec,R}).

Next, we turn to the loss probability $p$.
We show that for the case {\bf (i)}, i.e., if $0\leq R<1$,
then it is clear that $p<\pnec^{(0)}$ in (\ref{nec,3,R0}).
Since it holds that $R>\Rnec^{(0)}$, we have $\Rnec^{(0)}<1$.
Here, let $p^*$ denote the $p$ such that $\Rnec^{(0)}=1$ holds.
Since $\Rnec^{(0)}$ is monotonically increasing with respect to $p$,
$\Rnec^{(0)}<1$ implies that $p<p^*$.
On the other hand, after some calculation, we can show that
$p^*<\pnec^{(0)}$.
Thus,
the condition $p<\pnec^{(0)}$ is automatically satisfied.
Moreover, we can also show that $p^*<\pnec^{(1)}$.
Hence, $p<\pnec^{(1)}$ is necessary also in the case $0\leq R<1$.

Finally, since $\pnec^{(1)}$ must be larger than zero,
as a necessary condition, we need $0\leq\epsilon_n<1$ in (\ref{nec,p}).
\QED

\section{Sufficient Condition}\label{sufficient}
In this section, we present a sufficient condition for the existence of
a stabilizing feedback control scheme.
When we consider a practical quantization scheme,
we cannot choose the quantization level $N$ as a noninteger.
Therefore, we assume $N\in\Z$ and $N\geq 2$ in this section.

Given a certain data rate $R$, or $N$,
we employ the control law for the scaling parameter as
\begin{align}
 \sigma_k=&\mu(\calY^-_{k}),\label{suf,sigma}
\end{align}
and that for the control input as
\begin{align}
 u_k=&-\sum_{i=1}^n\left(a_i^*\hat y_{k-i+1}\right),\label{suf,u}
\end{align}
where
$\hat y_{k-i+1}:=(\widebar\calY_{k-i+1}+\wideubar\calY_{k-i+1})/2$.
Notice that we select the minimum scaling parameter  $\sigma_k$
satisfying (\ref{sigma_ineq}),
and take the state feedback control using the nominal values $a_i^*$ and
the centers of $\calY_{k-i+1}$.

For $i=1,2,\dots,n$, we introduce the following random variables $\theta_{i,k}$:
\begin{gather}
 \theta_{i,k}\!:=\!
 \begin{cases}
    |a_i^*|+\epsilon_i
    & \hspace{-.58em}\text{if }\gamma_{k-i+1}\!=\!0,\\
    \frac{|a_i^*|+\epsilon_i(N-1)}{N}
    & \hspace{-.58em}\text{if }\gamma_{k-i+1}\!=\!1 \text{ and } \calA_i\!\not\ni\! 0,\\
    \max\left\{\frac{|a_i^*|+\epsilon_i}{N},\epsilon_i\right\}
    & \hspace{-.58em}\text{if }\gamma_{k-i+1}\!=\!1 \text{ and } \calA_i\!\ni\! 0.
 \end{cases}\label{def,theta}
\end{gather}
This can be used to bound the interval $\calA_i\calY_{k-i+1}$ as
\begin{align}
 \mu(\calA_i\calY_{k-i+1})\leq\theta_{i,k}\sigma_{k-i+1}.\notag
\end{align}
Moreover,
define the random variable matrix $H_{\Gamma_k}$
containing the random variables $\theta_{1,k},\dots,\theta_{n,k}$ by
\begin{gather}
 H_{\Gamma_k}:=\left[\begin{array}{cccc}
   0& 1& \cdots& 0\\
   \vdots& \ddots& \ddots& \vdots\\
   0& 0& \cdots& 1\\
   \theta_{n,k} & \theta_{n-1,k} & \cdots&\theta_{1,k} 
 \end{array}\right],\notag
\end{gather}
where $\Gamma_k:=\{\gamma_{k},\gamma_{k-1},\dots,\gamma_{k-n+1}\}$.
Here, the process $\Gamma_k$ is a Markov chain
which has $2^n$ states given by
\begin{align}
 \Gamma^{(1)}:=\{0,\dots,0,0\},\
 \Gamma^{(2)}:=\{0,\dots,0,1\},\
 \dots,\
 \Gamma^{(2^n)}:=\{1,\dots,1,1\},\notag
\end{align}
and the transition probability matrix $P\!\in\!\R^{2^n\times2^n}$
is given by
\begin{align}
  P:=\left[
\begin{array}{cccccc}
 p& & & 1-p& & \\
 p& & & 1-p& & \\
  &\ddots& & &\ddots& \\
  & &p& & & 1-p\\
  & &p& & & 1-p
\end{array}\right],\notag
\end{align}
where the $(i,j)$ element is equal to the transition probability
from $\Gamma^{(i)}$ to $\Gamma^{(j)}$.
Furthermore, we define the matrix $F$ using $H_{\Gamma_k}$ and $P$:
\begin{gather}
 F:=F_1F_2,\notag
\end{gather}
where
\begin{align}
 F_1:=P^T \otimes I_{n^2},\
 F_2:=
 \diag\left(H_{\Gamma^{(1)}}\otimes H_{\Gamma^{(1)}},\dots,
 H_{\Gamma^{(2^n)}}\otimes H_{\Gamma^{(2^n)}}\right).\notag
\end{align}
Here, $\diag(\cdot)$ denotes a block diagonal matrix and $\otimes$ is the Kronecker product.
The following theorem holds 
by applying results from the Markov jump linear systems theory \cite{Costa2005}.
\begin{theorem}\label{th,suf}
 Given the data rate $R=\log_2 N$ and the loss probability $p\in[0,1)$,
if
\begin{align}
 \rho(F)<1\label{suf_cond},
\end{align}
then under the control law using (\ref{suf,sigma}) and (\ref{suf,u}),
the system depicted in \fref{fig,system} is MSS,
where $\rho(\cdot)$ represents the spectral radius of a matrix.
\end{theorem}

\begin{remark}\label{rem,iff}
For the special case of scalar plants ($n=1$),
the condition (\ref{suf_cond}) is equal to the following conditions:
\begin{align}
 R> \Rnec^{(1)},\
 p<\pnec,\
 0\leq \epsilon_n<1.\notag
\end{align}
This shows that if the necessary condition in Theorem \ref{th,nec} holds
and the data rate satisfies $2^R=N\geq2$,
then the sufficient condition is also satisfied.
Thus, Theorems \ref{th,nec} and \ref{th,suf} are necessary and sufficient
for the case $n=1$.
\end{remark}

\begin{remark}
 In \cite{Phat2004} and \cite{Martins2006},
sufficient conditions for stabilization of uncertain plants via lossless channels
($p=0$) are given.
For the case $n=1$,
the condition in \cite{Phat2004} is
\begin{align}
 R> R_{\text{P}}:=\log_2\frac{|\lambdaP|-\epsilon_1(|\lambdaP|+\epsilon_1)}
 {1-\epsilon_1(2|\lambdaP|+2\epsilon_1+1)},\label{Phat}
\end{align}
and the one from \cite{Martins2006} becomes
\begin{align}
 R> R_{\text{M}}:=\log_2\frac{|\lambdaP|}{1-\epsilon_1}.\label{Martins}
\end{align}
By Remark \ref{rem,iff},
our sufficient condition (\ref{suf_cond}) on the data rate equals
$R>\Rnec^{(1)}$ for the case $n=1$.
It is easy to verify that $\Rnec^{(1)}<R_{\text{P}},R_{\text{M}}$.
Thus,
our sufficient condition is less conservative than (\ref{Phat}) and (\ref{Martins}).
For general order plants, it is difficult to compare Theorem \ref{th,suf}
with the bounds in \cite{Phat2004} and \cite{Martins2006}
because the types of plant uncertainties are different.
\end{remark}

{\it Proof of Theorem \ref{th,suf}:~}
We first prove that if $\E[\sigma_k^2]\to 0$ then
$\E[|y_k|^2]\to0$ as $k\to\infty$ under the control law.
By substituting (\ref{suf,u}) to (\ref{AR}),
we obtain
\begin{align}
 |y_{k+1}|
 &=\left|\sum_{i=1}^n \left(a_iy_{k-i+1}-a_i^*\hat y_{k-i+1}\right)\right|\notag\\
 &\leq\sum_{i=1}^n\big(|a_i-a_i^*||\hat y_{k-i+1}|
  +|a_i|\left|y_{k-i+1}-\hat y_{k-i+1}\right|\big)\label{suf,1}.
\end{align}
Regarding the first term in the far right-hand side,
we have 
\begin{align}
 |a_i-a_i^*|\leq\epsilon_i\label{suf,1a}
\end{align}
from the bound on the uncertainty.
Moreover, since $\hat y_{k-i+1}$ is the midpoint of $\calY_{k-i+1}$,
we have
\begin{align}
  |\hat y_{k-i+1}|\leq\left|1-\frac{1}{N}\right|\frac{\sigma_{k-i+1}}{2}.\label{suf,1b}
\end{align}
Similarly, the second term
$\left|y_{k-i+1}-\hat y_{k-i+1}\right|$,
which corresponds to the quantization error,
is bounded by $\sigma_{k-i+1}$ as
\begin{align}
  \left|y_{k-i+1}-\hat y_{k-i+1}\right|
 \leq\frac{\sigma_{k-i+1}}{2^{R\gamma_{k-i+1}+1}}
 \leq\frac{\sigma_{k-i+1}}{2}.\label{suf,1c}
\end{align}
Applying these inequalities (\ref{suf,1a})--(\ref{suf,1c}) to (\ref{suf,1}),
we have
\begin{align}
 |y_{k+1}|
 &\leq\sum_{i=1}^n\epsilon_i\left|1-\frac{1}{N}\right|\frac{\sigma_{k-i+1}}{2}
 +|a_i|\frac{\sigma_{k-i+1}}{2}\notag\\
 &=\sum_{i=1}^n d_i\sigma_{k-i+1},\label{suf,deltasigma}
\end{align}
where $d_i$ is defined as follows:
\begin{align}
 d_i:=\epsilon_i\left|1-\frac{1}{N}\right|\frac{1}{2}+\frac{|a_i|}{2}
 \text{ for } i=1,2,\dots,n.\notag
\end{align}
By squaring both sides of the inequality in (\ref{suf,deltasigma})
and taking expectations,
we obtain
\begin{align}
 \E[|y_{k+1}|^2]
 &\leq\sum_{i=1}^n\sum_{j=1}^n d_i d_j
  \E\left[\sigma_{k-i+1}\sigma_{k-j+1}\right]\notag\\
 &\leq\sum_{i=1}^n\sum_{j=1}^n d_i d_j
  \sqrt{\E[\sigma_{k-i+1}^2]}\sqrt{\E[\sigma_{k-j+1}^2]}.\notag
\end{align}
Here, the second inequality holds by the fact that $\sigma_k>0$ for all $k\geq 0$
and the Schwarz inequality.
Hence,
we have that if $\E[\sigma_k^2]\to0$ then $\E[|y_{k+1}|^2]\to0$ as $k\to\infty$.

Next, we prove that the condition (\ref{suf_cond}) implies that
$\sigma_k$ is MSS under (\ref{suf,sigma}) and (\ref{suf,u}).
By (\ref{suf,sigma}) and the equality in (\ref{BM}), we have
\begin{align}
 \sigma_{k+1}=\sum^{n}_{i=1}\mu\left(\calA_i\calY_{k-i+1}\right).\label{suf,sigmamu}
\end{align}
For the $i$th term $\mu\left(\calA_i\calY_{k-i+1}\right)$,
it holds that
\begin{align}
\mu\left(\calA_i\calY_{k-i+1}\right)
 =\begin{cases}
    \left(|a_i^*|+\epsilon_i\right)\mu(\calY_{k-i+1})
    & \text{if }\calY_{k-i+1}\ni 0,\\
    |a_i^*|\mu(\calY_{k-i+1})+\epsilon|\widebar\calY_{k-i+1}+\wideubar\calY_{k-i+1}|
    & \text{if }\calY_{k-i+1}\not\ni 0 \text{ and } \calA_{i}\not\ni 0,\\
    2\epsilon_i\max\left\{|\widebar\calY_{k-i+1}|,\,|\wideubar\calY_{k-i+1}|\right\}
    & \text{if }\calY_{k-i+1}\not\ni 0 \text{ and } \calA_{i}\ni 0,
   \end{cases}\notag
\end{align}
by using basic results in interval arithmetics \cite{Moore1966} for $i=1,2,\dots,n$.
By taking the maximum of $\mu\left(\calA_i\calY_{k-i+1}\right)$
over all possible $\calY_{k-i+1}$, we have
\begin{align}
 \mu\left(\calA_i\calY_{k-i+1}\right)\leq\theta_{i,k}\sigma_{k-i+1},\notag
\end{align}
where $\theta_{i,k}$ is given in (\ref{def,theta}).
Thus, from (\ref{suf,sigmamu}), it follows that
\begin{align}
 \sigma_{k+1}
 \leq\theta_{1,k}\sigma_k+\theta_{2,k}\sigma_{k-1}+\cdots+\theta_{n,k}\sigma_{k-n+1}.\notag
\end{align}
This can be expressed as 
\begin{align}
  \zeta_{k+1}\leq H_k\zeta_k,\notag
\end{align}
where $\zeta_k:=[\sigma_{k-n+1}\;\cdots\;\sigma_k]^T$
and the inequality is in the element-wise sense.
From \cite[Theorem 3.9]{Costa2005},
it follows that the inequality (\ref{suf_cond}) implies that
$\zeta_k$ and hence $\sigma_k$ are MSS. 
\QED

We now illustrate the theoretical bounds on the data rate
obtained in Theorems \ref{th,nec}
and \ref{th,suf} by a numerical example.
Consider an uncertain plant of second order,
where $a_1^*=1$, $\epsilon_1=0.05$, $\epsilon_2=0.05$,
and $p=0.05$.
\begin{figure}[t]
\begin{center}
\includegraphics[scale=.45]{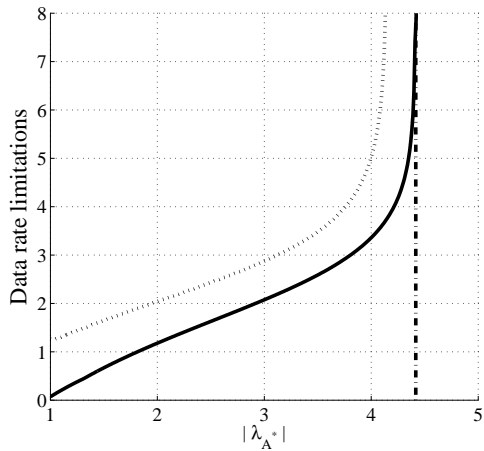}
\caption{Data rate limitations: The necessary condition (solid), the sufficient condition (dotted),
and the $\sup|\lambdaP|$ s.t.\ the loss probability $p=0.05$ satisfies $p<\pnec$ (dash-dot)}
\label{fig,plot}
\end{center}
\end{figure}
In \fref{fig,plot},
we plot the theoretical bounds on the data rate $R$
versus $|\lambdaP|=|a_2^*|$.
The vertical dash-dot line represents the supremum of $|\lambdaP|$
such that $p<\pnec$ holds.
Hence, the necessary condition (\ref{nec,p}) does not hold
on the right side of this line.
The figure shows a certain gap between the necessary condition (solid line) and
the sufficient condition (dotted line).
However, the gap is fairly small when $|\lambdaP|$ is sufficiently smaller
than the critical vertical line and is about one bit.

\section{Limitation on the Average Data Rate}\label{average}
In this section, we introduce a time-sharing protocol,
and then study conditions on the {\it average} data rate to stabilize the system.
So far, we have considered that the quantization level $N$ must be taken as an integer.
Hence, for a data rate $R$ satisfying the sufficient condition in Theorem \ref{th,suf},
the minimum quantization level is given by $N=\lceil 2^R \rceil$,
where $\lceil\cdot\rceil$ denotes the ceiling function.
Thus, the quantization level may be larger than $2^R$ due to the integer-constraint.
In a time-sharing protocol,
the observation of the plant output is $m$-periodic, where $m\in\N$ is the duration of
the time cycles.
Hence, by employing the protocol we can choose the quantization level
$N^m=\lceil 2^R \rceil$ and hence a noninteger $N$ as the average quantization level.

It is known that when we know the exact plant model $(A,B,C)$,
by taking the duration $m$ large enough, we can achieve the necessary bound in Theorem \ref{th,nec}.
That is, for any data rate greater than the bound, there exists a control law
with a feasible quantization level \cite{You2010, Tatikonda2004, Nair2004}.

Here, we will see that also when the plant is uncertain,
increasing the duration $m$ may help to reduce the required data rate.
However, for the duration above a certain value, we cannot stabilize
the system even with infinitely large data rate.
This is because a long duration, i.e., low observation frequency,
causes the accumulation of error in the state estimation due to the uncertainty.

To simplify the discussion,
in this section we consider the scalar plant given by
\begin{gather}
 y_{k+1}-a_ky_k=u_k,\label{AR1}\\
a_k\in\calA:=\left[a^*-\epsilon,a^*+\epsilon\right],\;
\epsilon\geq0,\;|a^*|-\epsilon>1.\notag
\end{gather}
Here, we introduce the encoding and decoding rules
based on a time-sharing protocol.
The basic idea is as follows:
Divide the time into cycles of a constant duration $m\in\N$
as $\left\{mj,mj+1,\dots,m(j+1)-1\right\}$, where $j\in\Zp$.
The encoder observes the output $y_{mj}$ at time $mj$,
and sends it to the decoder during the cycle from time $mj$ to $m(j+1)-1$.
Over the lossy channel,
the decoder receives $R\sum^{m-1}_{i=0}\gamma_{mj+i}$ bits in total in this cycle,
and compute the estimation set $\calY_{mj}$ of $y_{mj}$ at time $m(j+1)-1$.

More specifically, the encoder and the decoder function as follows.
At time $mj$,
the encoder quantizes $y_{mj}$ with the quantization level $N^m$.
We use the uniform quantizer as in Section \ref{formulation}.
The quantized signal is divided into packets of $R$ bits,
and is then transmitted to the decoder.
If a packet drops out then the encoder retries to send the same packet.
Notice that the encoder has access to the loss states
because of acknowledgement signals from the decoder.

The scaling parameter $\sigma_k$ is taken to be large enough to cover all possible outputs
at the next step.
In this section, since the observation is $m$-periodic,
we slightly change (\ref{sigma_ineq}) to
\begin{align}
 \sigma_{m(j+1)}\geq \mu(\calY_{m(j+1)}^-),\label{sigma_ineq2}
\end{align}
where $\calY_{m(j+1)}^-$ is the one cycle prediction set
based on the estimation set $\calY_{mj}$,
and is defined as follows:
\begin{align}
 \calY_{m(j+1)}^-:=\left\{(a')^m y'_{mj}+(a')^{m-1}u_{mj}+(a')^{m-2}u_{mj+1}
 +\cdots+a'u_{m(j+1)-2}
:a'\in\calA,\,y'_{mj}\in\calY_{mj}\right\}.\label{def,calY^-2}
\end{align}
Note that we have to take account of
the control inputs $u_k$ from $k=mj$ to  $m(j+1)-2$
since they affect the length of the prediction set.

We now study a condition for the stabilization of the system
under the protocol described above.

As we have shown in the proof of Theorem \ref{th,nec},
for $\lim_{k\to\infty}\E[|y_k|^2]=0$,
it is necessary that $\lim_{k\to\infty}\E[\sigma_k^2]=0$.
To derive a condition for stabilizing $\sigma_k$,
we follow an approach similar to that in Section \ref{necessary}.
In view of (\ref{sigma_ineq2}), 
we evaluate the length of the prediction set $\calY_{m(j+1)}^-$.
By (\ref{def,calY^-2}), we have
\begin{align}
 \mu(\calY_{m(j+1)}^-)
 &\geq\mu\left(\{(a')^m y'_{mj}:a'\in\calA,\,y'_{mj}\in\calY_{mj}\}\right)\label{m,1}\\
 &=
 \begin{cases}
  |a^*|^m\mu(\calY_{mj})+\delta_+|\widebar\calY_{mj}|+\delta_-|\wideubar\calY_{mj}| &
  \text{if } \calY_{mj}\not\ni0,\\
  \left(|a^*|^m+\delta_+\right)\mu(\calY_{mj}) &
  \text{if } \calY_{mj}\ni0,
 \end{cases}\notag
\end{align}
where $\delta_+$ and $\delta_-$ are nonnegative values defined as
$\delta_+:=(|a^*|+\epsilon)^m-|a^*|^m$ and $\delta_-:=-(|a^*|-\epsilon)^m+|a^*|^m$,
respectively.
The maximum of the right-hand side of the above equality with respect to $\calY_{mj}$
can be expressed as
\begin{align}
 \max_{\calY_{mj}}\mu\left(\{(a')^m y'_{mj}:a'\in\calA,\,y'_{mj}\in\calY_{mj}\}\right)
 =\begin{cases}
    \frac{|a^*|^m+\delta_+}{M_{mj}}\sigma_{mj} &
    \text{if }1\leq M_{mj}<2,\\
    \left(\frac{|a^*|^m-\delta_-}{M_{mj}}+\frac{\delta_++\delta_-}{2}\right)\sigma_{mj} &
    \text{if }M_{mj}\geq2,
  \end{cases}\label{m,2}
\end{align}
where $M_{mj}$ is given by
\begin{align}
 M_{mj}:=N^{\sum_{i=0}^{m-1}\gamma_{mj+i}}.\notag
\end{align}
Notice that the right-hand side of (\ref{m,2}) can be written as $\kappa_{mj}\sigma_{mj}$,
where the $\kappa_{mj}$ is the random variable defined as
\begin{align}
 \kappa_{mj}:=\frac{1}{M_{mj}}\biggl(|a^*|^m
 +\max\left\{\frac{M_{mj}}{2},1\right\}\delta_+
 +\max\left\{\frac{M_{mj}}{2}-1,0\right\}\delta_-\biggr).\notag
\end{align}
Thus, by (\ref{sigma_ineq2}), (\ref{m,1}), and (\ref{m,2})
it follows that
\begin{align}
 \sigma_{m(j+1)}\geq\kappa_{mj}\sigma_{mj}.\label{m,sigma_reccurence}
\end{align}
Since $\sigma_{mj}>0$ for all $j$,
and $\kappa_{mj}$ is independent of $\sigma_{mj}$,
$\sigma_k$ is MSS only if $\bar\kappa:=\E[\kappa_{mj}^2]<1$.
Thus, we have arrived at a necessary condition for the overall system to be MSS.
Note that the expectation $\bar\kappa$ is time-invariant
because it depends only on the process $\{\gamma_k\}_{k=0}^\infty$,
which is i.i.d.

As a stabilizing control law,
we employ the following:
\begin{align}
 \sigma_{m(j+1)}&=\mu(\calY_{m(j+1)}^-),\label{m,suf,sigma}\\
 u_{mj+i}&=\begin{cases}
      0 & \text{if } i\in\{0,1,\dots,m-2\},\label{m,suf,u}\\
      -(a^*)^m\hat y_{mj} & \text{if } i=m-1,
     \end{cases}
\end{align}
where $\hat y_{mj}:=(\widebar\calY_{mj}+\wideubar\calY_{mj})/2$.

The proposition below presents a necessary and sufficient condition
for the stabilization of the system employing the time-sharing protocol.
\begin{proposition}\label{prop,m}
 Consider the system depicted in \fref{fig,system}
with the uncertain scalar plant (\ref{AR1}), and a control law
under the time-sharing protocol.
If the system is MSS, then the following inequality must hold:
\begin{align}
 \bar\kappa<1.\label{m,cond}
\end{align}
Conversely,
if (\ref{m,cond}) holds
then the system is MSS under the control law (\ref{m,suf,sigma}) and (\ref{m,suf,u}).
\end{proposition}


{\it Proof:~}
(Necessity) Refer to the discussion before the proposition.

(Sufficiency)
As in the proof of Theorem \ref{th,suf},
we have that if $\lim_{k\to\infty}\E[\sigma_k^2]=0$ then
$\lim_{k\to\infty}\E[|y_k|^2]=0$
under the control law (\ref{m,suf,sigma}) and (\ref{m,suf,u}).
Moreover,
equality holds in (\ref{m,1}) and hence in (\ref{m,sigma_reccurence}).
Thus, for any $N$ and $m$ that satisfy (\ref{m,cond}),
$N^m\in\Z$, and $N^m\geq 2$,
we have that $\sigma_k$ is MSS.
\QED

From Proposition \ref{prop,m}, we can obtain the explicit limitation
on the data rate when the channel is lossless, i.e., for the case $p=0$.
To describe the limitation, let
\begin{align}
 \bRnec^{(1)}&:=\frac{1}{m}\log_2\frac{|a^*|^m-\delta_-}{1-(\delta_++\delta_-)/2}.\notag
\end{align}
Then, the condition (\ref{m,cond}) is equivalent to
\begin{align}
 R&>\bRnec:=\max\left\{\Rnec^{(0)},\bRnec^{(1)}\right\},\
 0\leq\frac{\delta_++\delta_-}{2}<1.\label{m,p=0}
\end{align}
Here, $\Rnec^{(0)}$ is the necessary bound defined in Section \ref{necessary}.
Clearly, the bound on the average data rate depends on the duration $m$,
and is equal to $\Rnec$ if $m=1$.
In particular, we cannot stabilize the system
even with large $R$ when $m$ is greater than a certain value;
if $m$ is large enough to satisfy $(\delta_++\delta_-)/2\geq1$
then $\bRnec^{(1)}=\infty$.

We illustrate the limitation on the average quantization level
given in Proposition \ref{prop,m}.
Consider an uncertain scalar plant, where $a^*=3.3$ and $\epsilon=0.025$
with a lossless channel $p=0$.
In \fref{fig,mVsN},
we plot the limitations on the average quantization level $N$
versus the duration $m$ of the cycles.
The vertical dash-dot line represents the supremum of $m$
such that the condition $(\delta_++\delta_-)/2<1$ in (\ref{m,p=0}) holds.
\begin{figure}[t]
\begin{center}
\includegraphics[scale=.5]{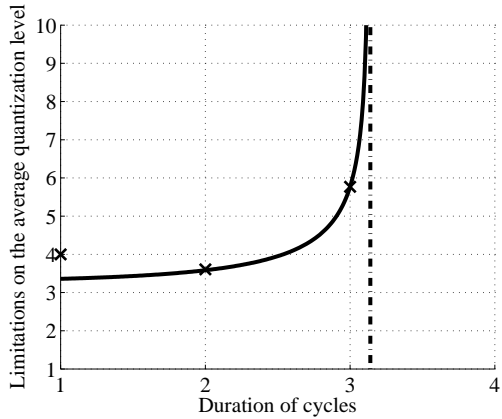}
\caption{Limitations on the average quantization levels:
The necessary bound (solid),
$\sup m$ s.t.\ $(\delta_++\delta_-)/2<1$ holds (dash-dot),
and the sufficient and feasible bound (cross)
}
\label{fig,mVsN}
\end{center}
\end{figure}
The necessary bound (solid line) increases with respect to the duration
and becomes infinitely large as it reaches the dash-dot line.
The cross marks represent
the sufficient and feasible average quantization levels,
i.e., the levels satisfying (\ref{m,p=0}), $N^m\in\Z$, and $N^m\geq2$.
We can find that the necessary bound on $N$ takes its minimum at $m=1$.
However, as the cross marks show,
when we take account of the feasibility
the duration $m=2$ is the best.

\section{Conclusion}\label{conclusion}
In this paper, we have studied the stabilization problem of uncertain networked control systems
under the presence of data rate constraints and packet losses.
In particular, we have derived a necessary condition and a sufficient condition
for the stability of the closed-loop system.
These conditions highlight limitations on the data rate, the loss probability,
and the uncertainty bounds,
and generalize existing results on nominal plants to the case with parametric uncertainties.


\end{document}